\documentclass[aps,pra,twocolumn,superscriptaddress,notitlepage,nofootinbib,longbibliography]{revtex4-2}
\usepackage{mathrsfs}
\usepackage{epsfig}
\usepackage{graphicx}
\usepackage{amsfonts}
\usepackage[figuresright]{rotating}
\usepackage{amssymb}
\usepackage{amsmath}
\usepackage{dcolumn}
\usepackage{bm}
\usepackage{xcolor}
\usepackage{braket}
\usepackage{units}
\usepackage{xspace}

\usepackage{bbold}
\definecolor{mypink3}{cmyk}{0, 0.7808, 0.4429, 0.1412}
\definecolor{mypink1}{rgb}{0.858, 0.188, 0.478}
\definecolor{mypink2}{RGB}{219, 48, 122}
\usepackage[colorlinks=true, allcolors=mypink1]{hyperref}
\usepackage[fleqn]{mathtools}

\usepackage[shortlabels]{enumitem}
\newcommand{\GCU}{Department of Physics, Government College University, Allama Iqbal, Faisalabad 38000, Pakistan}
\newcommand{\ZJNU}{Department of Physics, Zhejiang Normal University, Jinhua 321004, China}
\newcommand{\ECNU}{State Key Laboratory of Precision Spectroscopy, Quantum Institute for Light and Atoms, Department of Physics, East China Normal University, Shanghai 200062, China}
\newcommand{\NANKI}{The Key Laboratory of Weak-Light Nonlinear Photonics, Ministry of Education, School of Physics and TEDA Applied Physics Institute, Nankai University, Tianjin 300071, China}
\newcommand{\COMSATS}{Quantum Optics Lab, Department of Physics, COMSATS University, Islamabad 45550, Pakistan}
\begin{document}
\title{A rotational-cavity optomechanical system with two revolving cavity mirrors: optical response and fast-slow light mechanism}

\author{Amjad Sohail}
\affiliation{\GCU}
\author{Rameesa Arif}
\affiliation{\GCU}
\author{Naeem Akhtar}
\email{naeem\_abbasi@zjnu.edu.cn (First corresponding author)}
\affiliation{\ZJNU}
\author{Ziauddin}
\affiliation{\COMSATS}
\author{Jia-Xin Peng}
\email{18217696127@163.com (Second corresponding author)}
\affiliation{\ECNU}
\author{Gao Xianlong}
\affiliation{\ZJNU}
\author{ZhiDong Gu}
\affiliation{\NANKI}
\date{\today}

\begin{abstract}
We investigate the optical behavior of a single Laguerre-Gaussian cavity optomechanical system consisting of two mechanically rotating mirrors. We explore the effects of various physical parameters on the double optomechanically induced transparency (OMIT) of the system and provide a detailed explanation of the underlying physical mechanism. We show that the momentum is not the cause of the current double-OMIT phenomena; rather, it results from the orbital angular momentum between the optical field and the rotating mirrors. Additionally, the double-OMIT is simply produced using a single Laguerre-Gaussian cavity optomechanical system rather than by integrating many subsystems or adding the atomic medium as in earlier studies. We also investigate the impact of fast and slow light in this system. Finally, we show that the switching between ultrafast and ultraslow light can be realized by adjusting the angular momentum, which is a new source of regulating fast-slow light.
\end{abstract}
\maketitle
\section{Introduction}
Cavity optomechanics has focused a lot of interest on the investigation of coherent coupling between optical and mechanical modes via the radiation pressure of photons trapped inside an optical cavity \cite{1,Kip,Gir}. Additionally, in the contemporary era of quantum technology, such cavities have achieved significant advancements, including ultrahigh-precision measurement \cite{Teu,Naeem1}, gravitational wave detection \cite{Arva}, quantum information processing \cite{Tomb}, quantum teleportation \cite{Asjad1,Asjad2}, nonclassical photon statistics \cite{Raymond,Sithi,Sque}, higher-order sideband generation \cite{Xio,Caoo,Qian}, optomagnonic frequency combs \cite{comb}, quantum entanglement \cite{Sebe,Tian,Qubit}, macroscopic quantum coherence \cite{Spin,Whisp,Yao}, ground state cooling of mechanical oscillator \cite{6,Asjad}, and optomechanically induced transparency (OMIT) \cite{Dual,15,16,17}.

The electromagnetically induced transparency (EIT) is a quantum interference effect that has been shown in several quantum systems, including nitrogen vacancy centers \cite{11}, quantum dots \cite{12}, metamaterial \cite{PRB}, and Bose Einstein condensate \cite{14}. The OMIT phenomenon in the cavity optomechanical system was first reported theoretically \cite{15} before being demonstrated experimentally in \cite{16}. The OMIT phenomena is similar to the extensively investigated EIT phenomenon in the atomic medium \cite{9,10} because the radiation pressure produced mechanical oscillations in OMIT significantly change the optical response of a weak probing field.  Moreover, single OMIT and double-OMIT transparency windows have both been investigated in various quantum optomechanical systems, such as hybrid optomechanical cavities \cite{19,20,21,22,Phys,Russ,Russ1,23} and atomic-media assisted optomechanical systems \cite{24,25,26,27}. OMIT also has significant applications in the generation of fast/slow light \cite{33,34,35,Saif,Saiff} including light storage \cite{20,39,40} and the precise measurement of a variety of physical quantities such as electrical charge \cite{30}, mass sensor \cite{Mass}, and environmental temperature \cite{31}.

Most of the aforementioned research on OMIT is related to a typical cavity optomechanical system, where  the radiation pressure causes the interaction of the mechanical mode with the cavity field. This optomechanical interaction is mainly due to the exchange of linear momentum  between the mechanical mode and the cavity field. In particular, Bhattacharya and Meystre proposed a rotational-cavity optomechanical system in 2007 in which a macroscopic rotating mirror is coupled to a Laguerre-Gaussian (L-G) cavity mode through the exchange of orbital angular momentum and studied the cooling of the rotating mirror due to the action of the radiation torque \cite{Bhat}. Subsequently, Liu et al. also accomplished the ground-state cooling of the rotating mirror in a double  L-G cavity with atomic ensemble \cite{Liuu}. These studies on the ground cooling of the rotating mirror have provided many opportunities to further investigate macroscopic quantum phenomena in rotational cavity optomechanical systems, including bipartite quantum entanglement \cite{Bhat,4,5,Aixi} and tripartite quantum entanglement \cite{Tri}, macroscopic quantum coherence \cite{Maza}, OMIT phenomena \cite{Penng,Peng,ROS,AbbS,XUyi,QINLIAO}, and its applications for the measurement of orbital angular momentum \cite{Penng,Pei,MAbbs}, higher order sideband generation \cite{Second,HaoXion} and phase-matching analysis \cite{Phase}. However, compared to studies on conventional optomechanical systems, the research on the double-OMIT phenomenon in this type of rotating cavity optomechanical system is still essentially very limited. 

On the other hand, the realization of slow and fast light also makes the propagation of a weak probe field in cavity optomechanical systems extremely important \cite{M-D-Lukin-rmp, A-Kuzmich-prl}. The slow and fast light in atomic vapours and solid-state materials have been the subject of various investigations in the past. One application of these methods is to change the group velocity of the probe field, which results in the realization of either slow or fast propagation \cite{R-W-Boyd-sci}. For instance, slow light propagation occurs in atomic vaprs, the Bose-Einstein condensate \cite{A-Kasapi-prl, F-L-Kien-pra}, as well as in a medium of ultracold sodium vapours \cite{L-V-Hau-nat} and crystal \cite{M-S-Bigelow-sci}. Similar fast and slow light propagation in cavity optomechanical devices has also been demonstrated \cite{D-E-Chang}. Based on these previous studies, we are now interested to investigate the slow and fast light generation in the L-G rotational cavity optomechanical system.

In the present work, we investigate the double-OMIT phenomena in a single mode L-G cavity with two spiral phase elements serving as rotating mirrors, and a cavity field controlled by a strong pump field. The effects of many significant physical elements on OMIT have been thoroughly investigated, and thorough physical justifications have been provided. In particular, our approach opens up a brand-new opportunity to achieve double-OMIT in a single mode L-G cavity optomechanical system without the addition of any atomic medium or subsystem. Further,  we turn to the study of group delay in the system, which can be regarded as an analog of the fast-slow light in Fabry-P\'erot cavity.  The results show that switching between fast and slow light can be achieved by adjusting some parameters, such as orbital angular momentum. In addition, we found that the group delay can even reach $-400$ $\mu$s and $+400$ $\mu$s, which is a typical ultrafast-ultraslow light effect.

The paper is organized as follow. In Section II, we describe the model and introduce the Hamiltonian of the system. We also derive the quantum Langevin equations and provide the temporal evolution of the operator mean values in the same section. In Section III, we obtain the expressions of the output
 field at the probe frequency. The numerical results of the output field with various input field power, cavity decay rate, mechanical quality factor, and orbital angular momentum of the coupling field are discussed in Section IV. In Section V, we study the fast and slow light effects related to our system. Finally, we conclude our discussion in section VI.
 \begin{figure}
	\centering
	\includegraphics[width=0.5\textwidth]{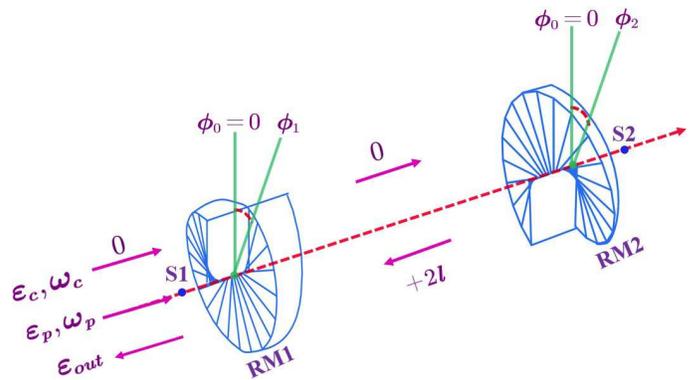}
	\caption{Schematic diagram of the L-G cavity optomechanical with two rotating mirrors, where a strong coupling field with zero topological charge at frequency $\protect\omega _{c}$\ and a weak probe field at frequency $\protect\omega _{p}$\ are input into the rotating cavity, they interact with two rotating mirrors at angular frequency $\protect\omega _{\protect\phi _{1}}$ and $\protect\omega _{\protect\phi _{2}}$, respectively. Besides, the angular displacement RM1 and RM2 are indicated by the angles $\phi _{1}$ and $\phi _{2}$  under the optorotational  interaction, respectively.} 	\label{Fig1}
	\label{Fig1}
\end{figure}
\section{Hamiltonian description}
\label{sec:model}
In this section, we present the model and introduce the Hamiltonian of our system. The system we considered is a rotating optomechanical cavity with length
$\mathcal{L}$ which consists of two rotating mirrors \cite{Bhat,5}, as shown in Fig.  (\ref{Fig1}).
The rotating mirror 1 (RM1) is partially transparent and the rotating mirror
2 (RM2) is perfectly reflecting. The field of a rotating cavity with frequency $%
\omega _{0}$\ is driven by a strong coupling field with amplitude $%
\varepsilon _{c}$\ at frequency $\omega _{c}$. Meanwhile, a weak probe field
with amplitude $\varepsilon _{p}$ and frequency $\omega _{p}$\ is used to detect the optical behavior of the system. We assume a strong coupling
field with zero topological charge (Gaussian beam) as an input to the cavity. The azimuthal
structure of the rotating mirrors removes the charges from the beam by
reflecting \cite{Bhat,Liuu,3,4,5}. When the coupling field is reflected by the RM1 and RM2, respectively, it acquires a topological charge of $-2l$ and $+2l$, demonstrating that the net topological charge increment of light transmission back and forth in the cavity is zero, ensuring the stability of the system. And this makes the strong
coupling field which couples the two rotating mirrors by the exchange of the
orbital angular momentum. Here we assume that the two rotating mirrors have the same
effective mass $m$ and radius $R$. Note that $\omega _{\phi _{j}} (j=1,2)$ and $\gamma _{j} (j=1,2)$ denote the angular frequency and intrinsic damping rate RM${j} (j=1,2)$, respectively.
 In a frame rotating at the driving frequency $%
\omega _{c}$, the Hamiltonian of the whole system takes the form
\begin{eqnarray}
\label{e1}
H/\hbar &=& \Delta _{0}a^{\dagger }a+\sum_{j=1,2} \dfrac{\omega _{\phi
		_{j}}}{2}( \phi _{j}^{2}+L_{z_{j}}^{2})+ (g_{1}\phi _{_{1}}-g_{2}\phi _{_{2}})a^{\dagger }a  \notag \\
&&+i \varepsilon _{c}(a^{\dagger }-a)+i\varepsilon _{p} \left( a^{\dagger }e^{-i\Delta t}-ae^{i\Delta t}\right) ,
\end{eqnarray}%
where $\Delta _{0}=\omega _{0}-\omega _{c}$  ($\Delta =\omega _{p}-\omega
_{c}$) is the detuning of the cavity field (the probe field) from the strong coupling field, $a$\ and $%
a^{\dagger }$\ ($\left[ a,a^{\dagger }\right] =1$) are the annihilation and
creation operators of the cavity field, respectively. And $\phi _{j}$ and $L_{z_{j}}$ with $[\phi _{j},L_{z_{k}}]=i\delta
_{jk}(j,k=1,2)$ are representing the angular displacement and angular momentum of the rotating mirrors, respectively. In Eq. \eqref{e1}, the first two terms are the free Hamiltonian of
the energy for the cavity field and the two rotating mirrors, the last two
terms give the coupling field and the probe field inside the cavity, the
third term describes the optomechanical coupling between the cavity field
and the two rotating mirrors with the coupling constant $g_{i}=(cL/\mathcal{L})\sqrt{%
\hbar /I\omega _{\phi _{i}}}$  \cite{Bhat}, where $c$ is the
velocity of light and $L$ is the orbital angular momentum quantum number, $\mathcal{L}$ is the length of the cavity and $I=mR^{2}/2$ is the moment of inertia of the two rotating mirrors about the central axis of the cavity. In addition, the coupling field amplitude and the probe
field are $\varepsilon _{c}=\sqrt{2\kappa P/\hbar \omega _{c}},$\ $%
\varepsilon _{p}=\sqrt{2\kappa P_{p}/\hbar \omega _{p}}$, respectively. Here, $P$ ($P_{p}$) and $\omega _{c}$ ($\omega _{p}$) represent the power and frequency of the coupling (probe) field.
\begin{figure*}
\centering
	\includegraphics[width=0.8\textwidth]{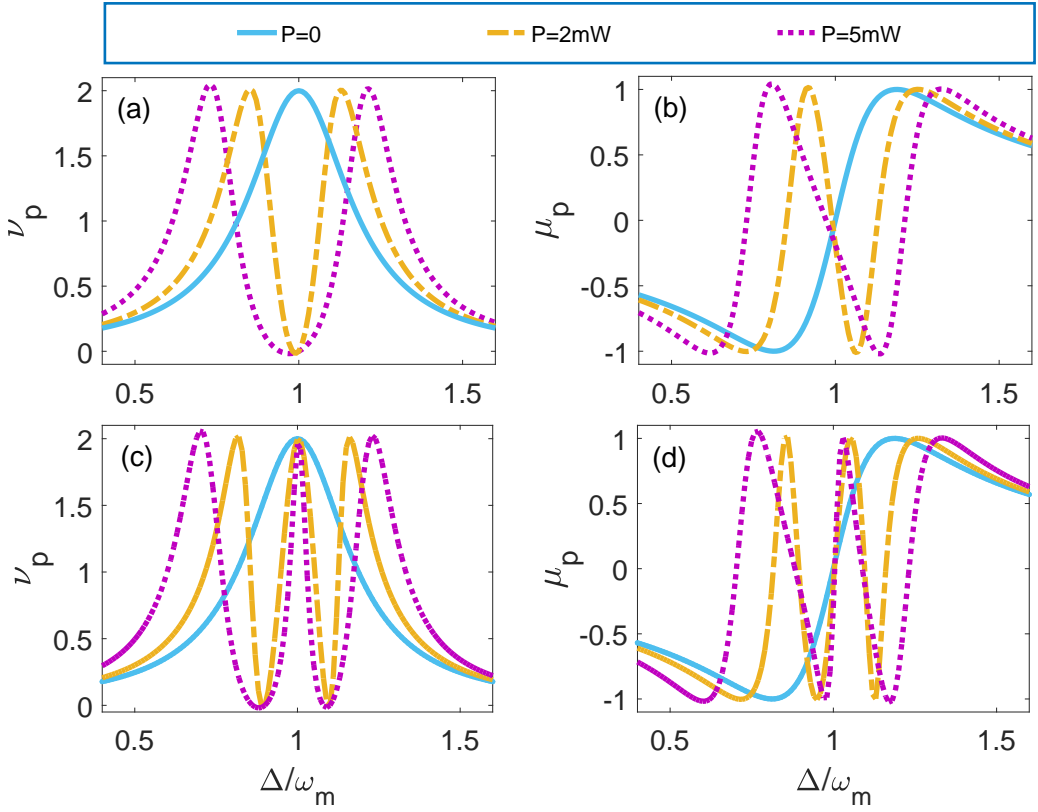}
	\caption{The absorption spectra $\protect\nu _{p}\ $of the
		output probe field as a function of normalized $\Delta \diagup \protect\omega _{m}$ for different power $P=0$, 2 mW and 5 mW, respectively. The angular frequency of rotating mirrors $\protect\omega _{\protect\phi _{1}}=\protect\omega _{\protect\phi _{2}}=\protect\omega _{m}$ in (a-b) and $\protect\omega _{\protect\phi _{1}}=1.1\protect\omega _{m}$, $\protect\omega _{\protect\phi _{2}}=0.9\protect\omega _{m}$ in (c-d).} \label{Fig2a}
\end{figure*}

On the basis of the aforementioned, we examine the mean value Heisenberg equation, which is derivable from the Hamiltonian in Eq. \eqref{e1}, while taking into account the dissipation of the cavity field and the damping of the revolving mirrors, and excluding quantum noise and thermal noise. In the strong driving
regime we have $\langle a\rangle \gg 1$, thus the
factorization assumption $\langle \phi _{i}a\rangle =\langle \phi
_{i}\rangle \langle a\rangle \left( i=1,2\right) $\ and $\langle a^{\dagger
}a\rangle =\langle a^{\dagger }\rangle \langle a\rangle $ are valid. This implies that the
time evolution of the mean value equation of the system operators gets form as
\begin{eqnarray}
\label{e2}
\langle \dot{\phi _{j}}\rangle &=&\omega _{_{\phi
		_{j}}}\langle L_{z_{j}}\rangle, \notag \\
\langle \dot{L_{z_{j}}}\rangle &=&-\omega _{_{\phi
		_{j}}}\langle \phi _{j}\rangle +g_{\alpha{j}}\langle a^{\dagger }\rangle \langle
a\rangle -\gamma _{j}\langle L_{z_{j}}\rangle, \\
\langle \dot{a}\rangle &=&-i[\Delta _{0}+g_{1}\phi
_{_{1}}-g_{2}\phi _{_{2}}]\langle a\rangle -\kappa \langle a\rangle
+\varepsilon _{c}+\varepsilon _{p}e^{-i\Delta t}  \notag,
\end{eqnarray}
where $ j=1, 2$,  $g_{\alpha{1}}=-g_{1}$ and $g_{\alpha{2}}=g_{2}$.
In the weak driving regime, especially in the single-photon strong coupling
regime, the response of the system to the probe field will be
affected by the nonlinear interaction between the cavity field and each
rotating mirror. Since the probe field is much weaker than
the coupling field $(\left\vert \varepsilon _{p}\right\vert \ll \varepsilon
_{c})$, the steady-state solution to Eqs. \eqref{e2} can be approximated to the
first order in the probe field $\varepsilon _{p}$. In the long time limit,
the solution to Eqs. \eqref{e2} can be written as \cite{Dual,15,16,17, ASR}
\begin{equation}
\label{e3}
\langle s\rangle =s_{0}+s_{+}\varepsilon _{p}e^{-i\Delta t}+s_{-}\varepsilon
_{p}^{\ast }e^{i\Delta t},
\end{equation}%
where $s=[\phi _{1},\phi _{2},L_{z_{1}},L_{z_{2}},a]$. Note that here we retained only the components $\omega _{c},$\ $\omega _{p}, $\ $
2\omega _{c}-\omega _{p}$. Substituting the Eq. \eqref{e3} into the
Eqs. \eqref{e2}, we can obtain the following expressions
\begin{eqnarray}
\label{e4}
L_{z_{j}} &=&0,\phi _{j0}=\frac{g_{\alpha {j}}\left\vert a_{0}\right\vert
	^{2}}{\omega _{\phi _{j}}},a_{0}=\frac{\varepsilon _{c}}{\kappa +i\Delta
	^{\prime }},  \nonumber \\
a_{+} &=&\frac{1}{d(\delta )}\left[ A\ast \Lambda _{1}\ast \Lambda
_{2}+i(G_{1}^{2}\omega _{\phi _{1}}\Lambda _{2}+G_{2}^{2}\omega _{\phi
	_{2}}\Lambda _{1})\right] ,  \nonumber \\
a_{-} &=&\frac{1}{d^{\ast }(\delta )}\left[ \frac{a_{0}^{2}}{\left\vert
	a_{0}\right\vert ^{2}}i(G_{1}^{2}\omega _{\phi _{1}}\Lambda
_{2}+G_{2}^{2}\omega _{\phi _{2}}\Lambda _{1})\right] , \\
d(\delta ) &=&A\ast A^{\prime }\ast \Lambda _{1}\ast \Lambda _{2}-2\Delta
^{\prime }(G_{1}^{2}\omega _{\phi _{1}}\Lambda _{2}+G_{2}^{2}\omega _{\phi
	_{2}}\Lambda _{1}).  \nonumber
\end{eqnarray}
Here $A=\kappa -i(\Delta ^{\prime }+\Delta ),A^{\prime }=\kappa +i(\Delta
^{\prime }-\Delta ),\Lambda _{j}=\omega _{\phi _{j}}^{2}-\Delta ^{2}-i\gamma
_{j}\Delta ,G_{j}=g_{\alpha {j}}\left\vert a_{0}\right\vert (j=1,2)$.
Among $\Delta ^{\prime }=\Delta _{0}+g_{1}\phi _{10}-g_{2}\phi _{20}$\ is
the detuning between the effective frequency of L-G cavity
and the frequency of the coupling field, and it depends on $a_{0}$.
Additionally, $G_{1}$ and $G_{2}$ can be considered of as, respectively, the effective optomechanical coupling rates to the cavity field and RM1 and RM2.

\section{The output field}
In this section, we solve the properties of the components of the optical field output
at the probe field frequency $\omega _{p}$. According to the input-output
relation, we can get the magnitude of the output field as follows \cite{Dual,15,16,17, ASR},
\begin{equation}
\label{e5}
\varepsilon _{out}(t)+\varepsilon _{c}+\varepsilon _{p}e^{-i\Delta
	t}=2\kappa \left\langle a\right\rangle .
\end{equation}
We write the output field $\varepsilon _{out}(t)$\ in a similar form to Eq.
\eqref{e3}, then expand the amplitude of the output field by a small amount of $%
\varepsilon _{p}$\ with
\begin{equation}
\label{e6}
\varepsilon _{out}(t)=\varepsilon _{out0}+\varepsilon _{out+}\varepsilon
_{p}e^{-i\Delta t}+\varepsilon _{out-}\varepsilon _{p}^{\ast }e^{i\Delta t},
\end{equation}%
where $\varepsilon _{out0}$, $\varepsilon _{out+}$, and $\varepsilon _{out-}$%
\ are the components of the output field at frequency $\omega _{c},$\ $%
\omega _{p},$\ $2\omega _{c}-\omega _{p}$, respectively.  Here $\varepsilon _{out-}$\ is
known as the Stokes field, and it is generated from the process where two
photons at angular frequency $\omega _{c}$ interact with a single photon at
frequency $\omega _{p}$\ to create a new photon at frequency $2\omega
_{c}-\omega _{p}$. Substituting the Eq. \eqref{e6}, the part of Eqs. \eqref{e4} and Eq.
\eqref{e3} into the Eq. \eqref{e5}, the output
field at the weak probe frequency is
\begin{eqnarray}
\label{e7}
\varepsilon _{out+} &=&2\kappa a_{+}-1.
\end{eqnarray}%
 The quadratures of the output field defined by
\begin{equation}
\varepsilon _{T}=\varepsilon _{out+}+1=2\kappa a_{+}=\nu _{p}+iu_{p}.
\end{equation}%
Here $\nu _{p}$\ and $u_{p}$\ are the two quadratures components of the
output field $\varepsilon _{T}$, where the real part $\nu _{p}$\ and the
imaginary part $u_{p}$\ represent the absorption characteristic and the
dispersion characteristic of the cavity optomechanical system, respectively. The homodyne technique can be used to measure the field quadratures, as is widely known \cite{15,16}.

\section{Analyzing the output probe field}

Let us now discuss the effect of various physical factors on the output field, such as the power of the pumping field, the mechanical frequencies of two rotating mirrors, angular quantum number of L-G cavity modes, and damping of the system.\;We utilize similar parameter values to those in the papers \cite{Bhat,Liuu,4,5,Aixi,Penng,Peng} for our numerical calculations. In particular, we choose the wavelength of the
strong coupling field $\lambda _{c}=810$ nm, the orbital angular momentum
number $L=100$, the masses and the radius of the rotating mirrors $m_{1}=m_{2}=m=50$ ng\
and $R=0.1$ $\mu$m, the L-G cavity mode decay rate $\kappa =15\pi \times
10^{6}$ Hz, the angular frequencies of the rotating mirrors $\omega _{\phi_{1}}=1.1 \omega _{m}$ and $\omega _{\phi _{2}}=0.9\omega _{m}$,
respectively, where $\omega _{m}=160\pi \times 10^{6}$ Hz. The
mechanical quality factors of the rotating mirrors are $Q_{1}=Q_{2}=Q=1.2\times
10^{5}$ and the mechanical damping rate is $\gamma _{i}=\omega _{\phi _{i}}/
Q_{i}(i=1,2)$. In particular, the system is
placed in the resolved side-band regime $(\kappa /\omega _{m}=0.187<1)$, which facilitates the study of anti-Stokes processes. Besides, the strong coupling field is tuned close to the red
side-band of the cavity resonance, thus the effective detuning approximately satisfies the relation $\Delta
^{\prime }=\omega _{m}$ \cite{Dual,15,16,17}.

\subsection{Pumping field versus the output probe field}

In Fig. (\ref{Fig2a}), we plot the absorption $\nu _{p}$ and dispersion $u_{p}$ of the output probe field as a
function of the normalized detuning $\Delta /\omega _{m}$ for
three different values of the power $P$ of the coupling field. We consider
both the rotating mirror RM1 and RM2 having the same angular frequency, i.e., $\protect\omega _{\protect\phi _{1}}=\protect\omega _{\protect\phi _{2}}=\protect\omega _{m}$. In Fig. \ref{Fig2a}(a), it is clear that the absorption curve exhibits a standard Lorentzian line pattern when $P=0$, which indicates that the probe field is strongly absorbed if $\Delta =0$. However, as the power $P$ increases, it is evident that the absorption peak transforms into a dip, as is typical for an EIT curve. In particular, the greater the power of the pump field results into the dip of the wider width and deeper depth, hence, causing the more significant the OMIT effect. When the power is not zero, the probe light exhibits a dramatic dispersion change at the dip position, as seen in Fig. \ref{Fig2a} (b). The later study of fast and slow light in this paper is built on this phenomena. It is interesting to note that two revolving mirrors, with angular frequencies of $\omega
_{\phi _{1}}=1.1\omega _{m}$\ and $\omega _{\phi _{2}}=0.9\omega _{m}$, have distinct fundamental frequencies. Figure \ref{Fig2a}(c) demonstrates that the absorption curve dips at $\Delta = 0.9\omega _{m}$ and $\Delta =1.1\omega _{m}$ when the pumping field power is not zero, which is a common double-OMIT phenomenon. The left transparent window is caused by the coupling of RM2 and the cavity field, while the right transparent window is caused by the coupling of RM1 and the cavity field.

\begin{figure}
\centering
\includegraphics[width=0.45\textwidth]{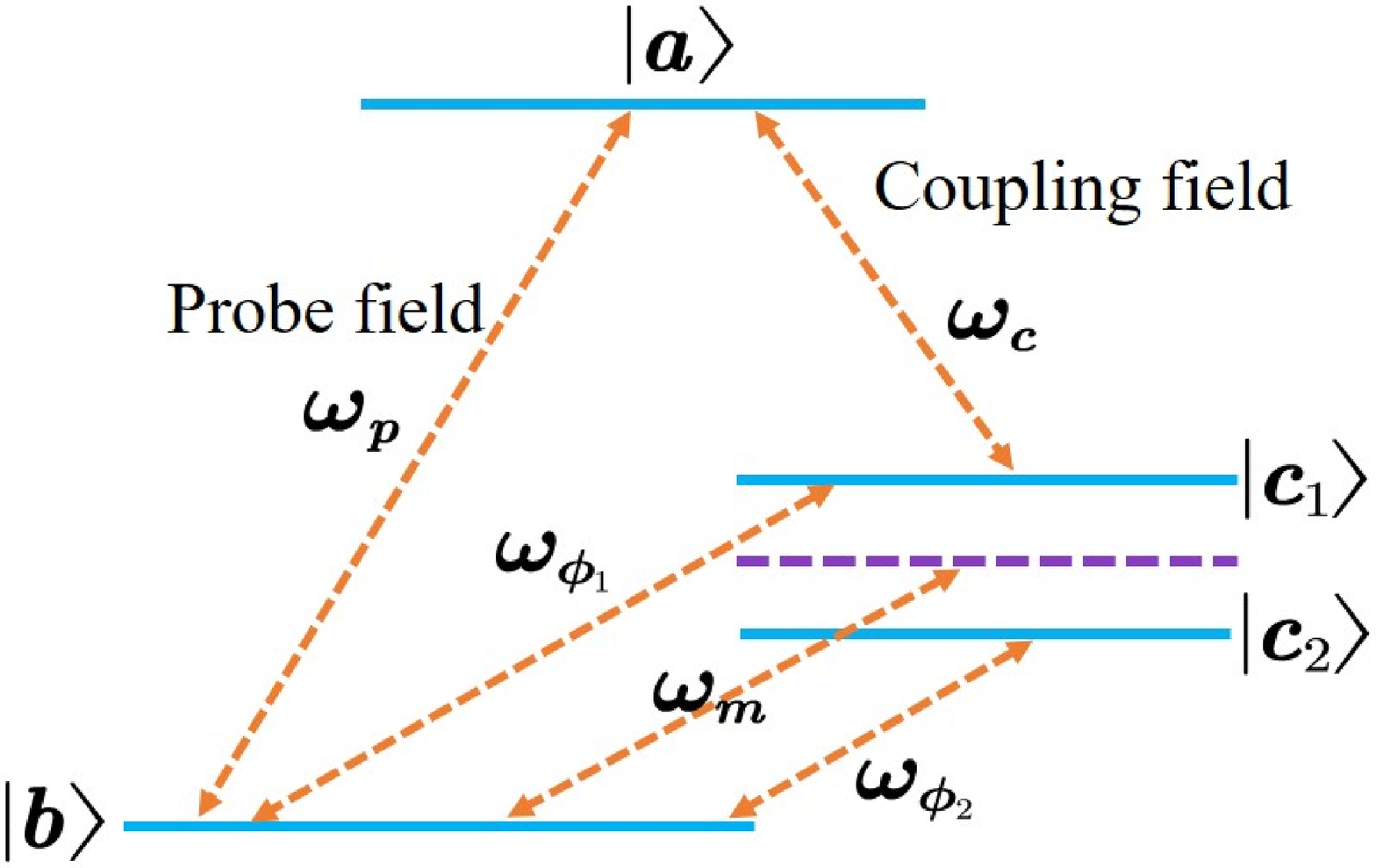}
	\caption{The cavity frequency of the excitation of $\left\vert a\right\rangle \leftrightarrow \left\vert b\right\rangle  $ and $\left\vert a\right\rangle \leftrightarrow \left\vert c_{i}\right\rangle (i=1,2)$ are $\omega _{p}$ and $\omega _{c}$, respectively. The frequency of the rotating mechanical
		excitation $\left\vert c_{1}\right\rangle \leftrightarrow \left\vert b\right\rangle $ is $\protect\omega _{\protect\phi _{1}}$ and $\left\vert c_{2}\right\rangle \leftrightarrow\left\vert b\right\rangle $ is $\protect\omega _{\protect\phi _{2}}$. $\omega _{m}=(\omega _{\phi _{1}}+\omega
		_{\phi _{2}})/ 2$ is normalized frequency.} \label{Fig2b}
\end{figure}

In particular, the two rotating mirrors in our system rotate about the cavity axis; as a result, their motion can be explained by rotational modes of the RM phonon states \cite{Bhat,Liuu}. By combining the photon state of the cavity field, they create the energy-level diagram shown in  Fig. (\ref{Fig2b}), which has $\Lambda$-type energy level structure. Therefore, the physical mechanism behind the above phenomena can be easily explained by energy level diagram of the system. In Fig. (\ref{Fig2b}), when $\protect\omega _{\protect\phi _{1}}=\protect\omega _{\protect\phi _{2}}=\protect\omega _{m}$, it is clear that the detuning $\Delta=\omega _{p}-\omega _{c}$ is equal to $\omega _{m}$, $\left\vert a\right\rangle \leftrightarrow \left\vert b\right\rangle  $ will have two different transition channels (one $\Lambda $-type energy level structure) to interfere. That is the anti-Stokes field
at frequency $\omega _{c}+\omega _{m}$ and the input probe field at
frequency $\omega _{p}$  cause the destructive interference, and then
it has one EIT dip at $\Delta =\omega _{m}$ on the absorptive curve of
output field, this is the standard model for appearing single transparent
window. However, when $\protect\omega _{\protect\phi _{1}}$ is not equal to $\protect\omega _{\protect\phi _{2}}$, there is obviously one more transition channel than before (two $\Lambda $-type energy level structures). Therefore, it leads to the double-OMIT phenomenon. In addition, the position of the transparent window is closely related to the frequency of the rotating mirror. It is important to note that, unlike in the previous studies, where the double-OMIT phenomena was caused by the integration of many subsystems or the addition of an atomic medium, it is now only the outcome of a single L-G cavity system. So the research model we currently use is simpler.
\begin{figure}
\centering
\includegraphics[width=0.5\textwidth]{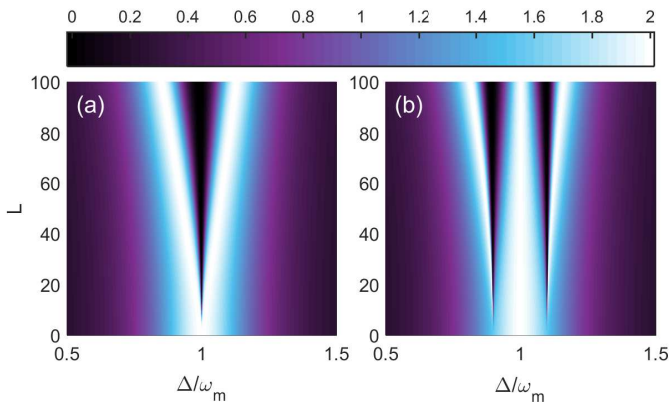}
\caption{The contour plot of absorption spectra $\protect\nu _{p}\ $ of the output probe field as a function of normalized $\Delta /\protect\omega _{m}$ and angular momenta $L$ when (a) $\protect\omega _{\protect\phi _{1}}=\protect\omega _{\protect\phi _{2}}=\protect\omega _{m}$ and (b) $\protect\omega _{\protect\phi _{1}}=1.1\protect\omega _{m}$, $\protect\omega _{\protect\phi _{2}}=0.9\protect\omega _{m}$.} \label{Fig4}
\end{figure}
\subsection{Orbital angular momentum of the coupling field}

In this subsection, we examine the effect of the orbital angular momentum of
the coupling field on the absorption of the output probe field. In Fig. (\ref{Fig4}) we have drawn diagrams of the relationship density
plot of absorption spectra versus scaled $\Delta /\protect\omega _{m}$ and angular momenta $L$. Similar to the previous results, when $\protect\omega _{\protect\phi _{1}}=\protect\omega _{\protect\phi _{2}}$, only one transparent window appears, while when $\protect\omega _{\protect\phi _{1}}$ is not equal to $\protect\omega _{\protect\phi _{2}}$, two transparent windows are displayed. Interestingly, we can clearly see that with the increase of angular momentum, the width and depth of the transparent window increases. This result is understood by the optorotational coupling strength $g_{i}=(cL_{i}/\mathcal{L})\sqrt{\hbar /I\omega _{\phi _{i}}}\left( i=1,2\right) $. According to this formula, we know that, when $L$ is small, the effective coupling strength between the cavity mode and the rotating mirrors will be weak, which inevitably lead to the weakening of the anti-Stokes process in the system. This make the destructive interference effect between the anti-Stokes field and the probe field less significant. Hence, as a result, the OMIT effect is not obvious, and vice versa. The current double-OMIT phenomenon originates from the exchange of orbital angular momentum between the light field and the rotating mirrors, rather than the momentum.

\subsection{Effects of dissipation on the output field}

Here, we explore the effect of the dissipation channels on the output probe field. For the currently studied optomechanical system,
the main dissipation channels are the thermal damping of the rotating mirrors and the cavity mode decay. Figure (\ref{Fig5}) exhibits the role of the damping of the rotating mirrors on the  absorption spectra. In Fig. \ref{Fig5}(a), we show that the mechanical quality factor of RM1 decreases (corresponding to increase the damping), and the depth of the right transparent window decreases. Moreover, here we also show that the transparent window on the left is almost invariant. This happen due to the transparent window on the right that is build by RM1, and thermal damping harmful effect on the destructive interference between the anti-Stokes field $\omega _{c}+\protect\omega _{\protect\phi _{1}}$ and the output probe field $\omega _{p}$. Evidently, the influence of damping of RM2 on the absorption spectra lead to identical results, as illustrated in Fig. \ref{Fig5}(b).

We also study the influence of cavity field attenuation on the absorption curve, and this is illustrated in Fig. (\ref{Fig6}). We find that the two transparency windows and the center peak have no change with the increase of decay rate $\kappa $, while sidebands on both sides of the transparent windows become wider (this can be considered as the spectral line broadening caused by cavity field attenuation). This phenomenon shows that the double-OMIT phenomenon is robust against cavity field attenuation. Physically, the increase of cavity field attenuation inevitably weaken the optomechanical coupling between the cavity field and rotating mirrors. More importantly, when the angular displacement of rotating mirrors is large then a large strain is needed to reduce the optorotational effect and vice versa. Hence, the current research system shows strong robustness to cavity field attenuation.

\section{Fast and slow light concept}

In the previous section, we have investigated the optical response in the L-G
cavity optomechanical system consist of two rotating mirrors. In particular, we
have shown that the dispersion curve shows a sharp dispersion change at the dip,
which prompted the study of fast-slow light. In general, the phase of
output probe field is related to the group delays of the output field and is
given by \cite{33,34,35,Saif,Saiff}
\begin{equation}
\label{e9}
\tau _{g}=\frac{d\phi \left( t_{p}\right) }{d\omega _{p}},
\end{equation}%
where $t_{p}=1-\varepsilon _{T}$ is transmitted field and $\phi \left(
t_{p}\right) =\arg \left[ t_{p}\right] $ represents the phase of  $t_{p}.$
 Note that $\tau _{g}>0$ (positive group delay) denotes slow light,
whereas $\tau _{g}<0$ (negative group delay) represents the fast light.
In the following, we first study the effects of physical parameters on the
phase dispersion $\phi \left( t_{p}\right) $, and then analyze their effects on group delay $\tau _{g}$.
\begin{figure}
\centering
\includegraphics[width=0.4\textwidth]{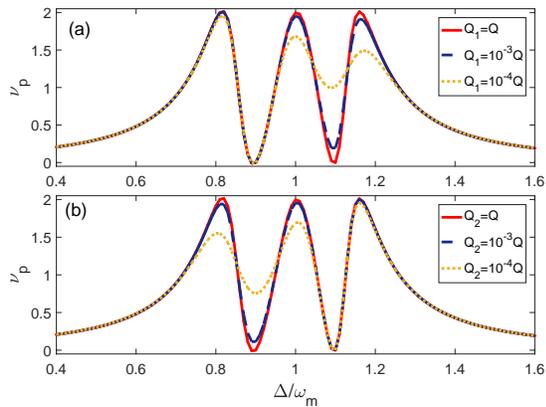}
\caption{The plot of absorption spectra $\protect\nu _{p}\ $ of the output probe field as a function of normalized $\Delta/\omega _{m}$\ for different values of mechanical quality factors. Here, we set $Q=1.2\times 10^{5}$.} \label{Fig5}
\end{figure}
\begin{figure}
\centering
\includegraphics[width=0.4\textwidth]{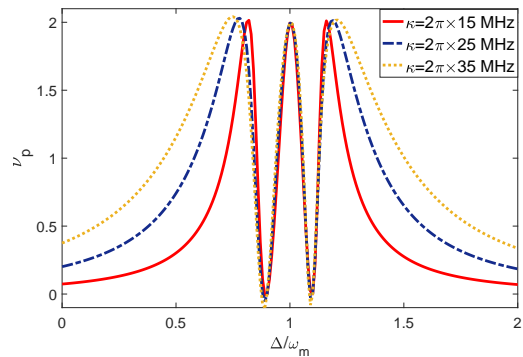}
\caption{The plot of absorption spectra $\protect\nu _{p}\ $ of the output probe field as a function of normalized $\Delta/\omega _{m}$ for different values of cavity decay rate.} \label{Fig6}
\end{figure}
\begin{figure*}
\centering
\includegraphics[width=0.45\textwidth]{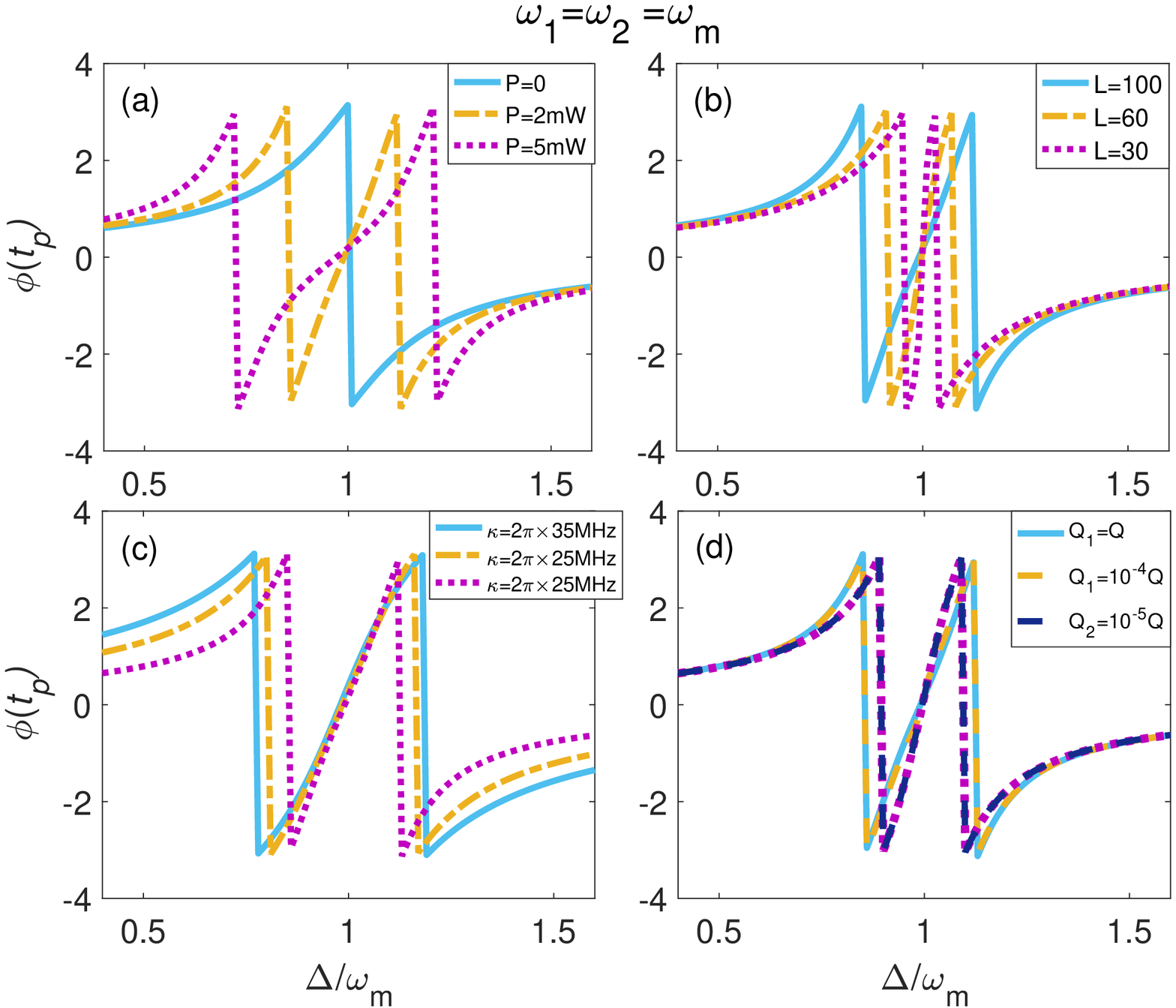}
\includegraphics[width=0.45\textwidth]{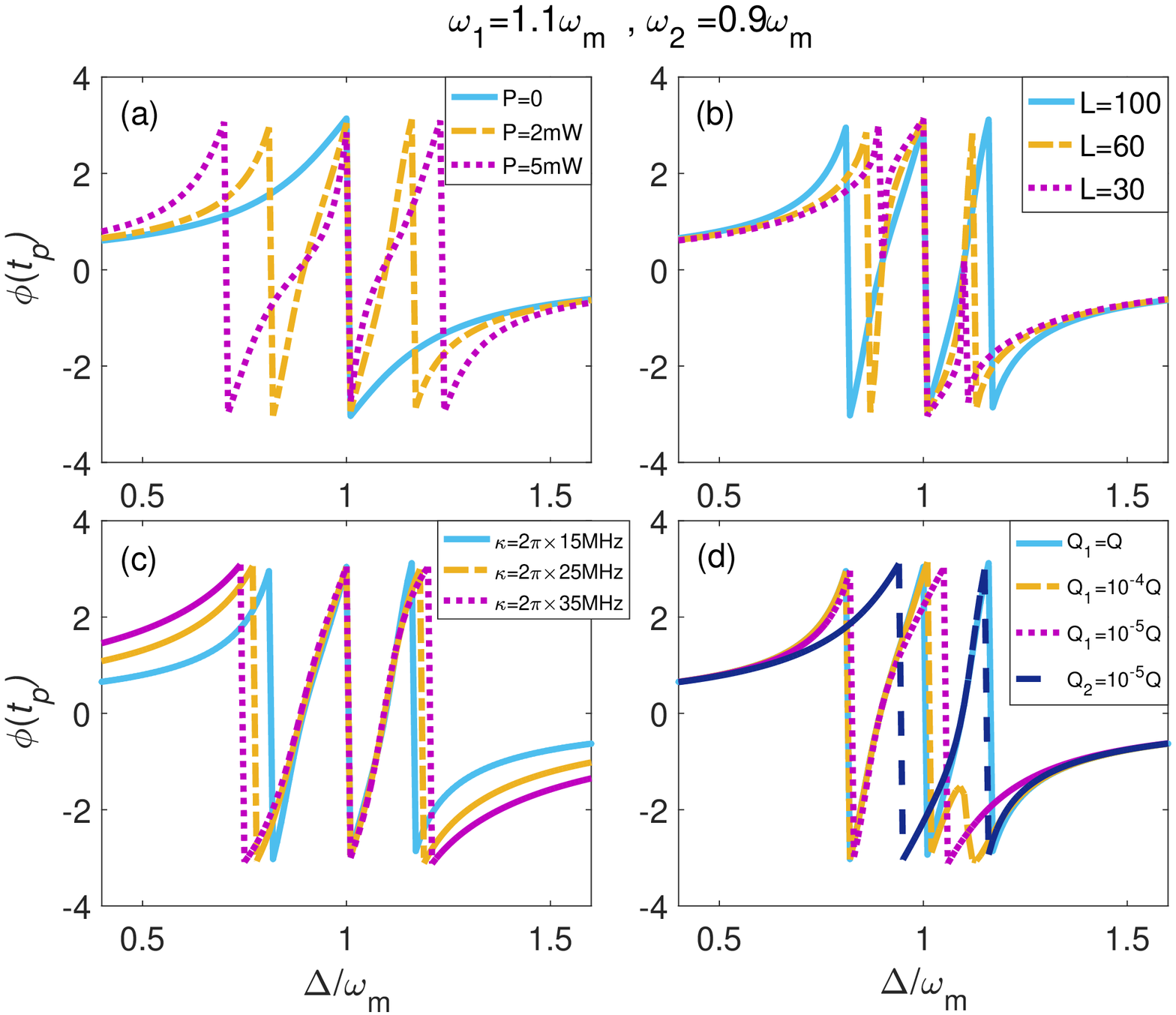}
\caption{The phase dispersion $\phi \left( t_{p}\right)$ as a function of normalized $\Delta/\omega _{m}$ for different the coupling field power, angular momentum quantum number, cavity field attenuation and mechanical quality factor. 	The four subgraphs above correspond to  $\protect\omega _{\protect\phi _{1}}=\protect\omega _{\protect\phi _{2}}$, while the four subgraphs below correspond to  $\protect\omega _{\protect\phi _{1}}=1.1\protect\omega _{m}$, $\protect\omega _{\protect\phi _{2}}=0.9\protect\omega _{m}$.
Here, we set $Q=1.2\times 10^{5}$, and the statement in the first paragraph of Section IV for other parameters.} \label{Fig7}
\end{figure*}
\begin{figure}
\centering
\includegraphics[width=0.5\textwidth]{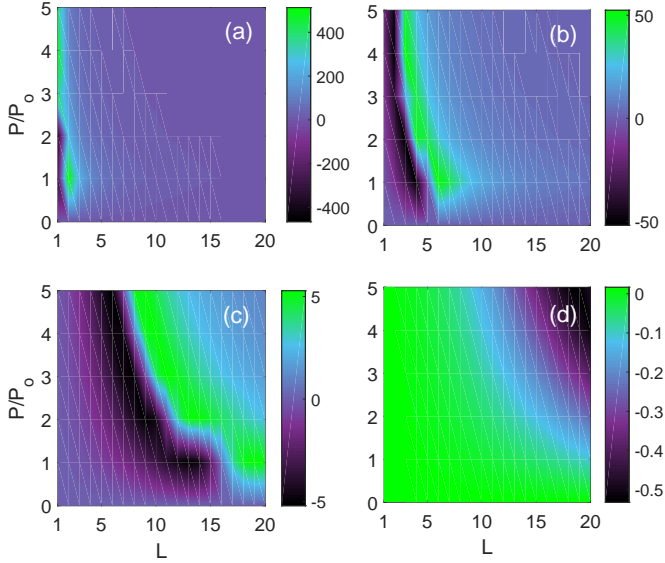}
\caption{The density plot of the group delay $\tau _{g}$ (in $\mu$s) versus the  coupling field power and the angular momentum quantum number for the different values of mechanical quality factors. We set  $\protect\omega _{\protect\phi _{1}}=1.1\protect\omega _{m}$, $\protect\omega _{\protect\phi _{2}}=0.9\protect\omega _{m}$, $P_{0}=1\times 10^{-3}$ mW, and $\Delta=1.1\omega _{m}$. (a) $Q_{1}=Q$, (b) $Q_{1}=10^{-1}Q$, (c) $Q_{1}=10^{-2}Q$, (d) $Q_{1}=10^{-3}Q$, here $Q_{2}=Q=1.2\times 10^{5}$. } \label{Fig8}
\end{figure}

Figure (\ref{Fig7}) is the phase dispersion $\phi \left( t_{p}\right) $ depending on the  normalized detuning $\Delta/\omega _{m}$ under different physical parameters, such as the  coupling power, the angular momentum quantum number, the cavity field attenuation and quality factor of rotating mirror. One can clearly see that when $\protect\omega _{\protect\phi _{1}}=\protect\omega _{\protect\phi _{2}}$, the phase dispersion $\phi \left( t_{p}\right) $ presents a zigzag shape with two dips as $\Delta=\omega _{m}$ changes. Because the derivative of $\phi \left( t_{p}\right) $ with respect to $\Delta$ is the same as that with respect to $\omega _{p}$. Equation \eqref{e9} reveals the emergence of fast and slow light. Physically, the sharp change of $\phi \left( t_{p}\right) $ results in a huge change in the refractive index of the medium, here causing group advanced and delay phenomena. In particular, we find that the distance between two dips change with the adjustment of physical parameters, which is much similar to the dispersion curve (see Fig. \ref{Fig2a}(b) and Fig. \ref{Fig2a}(d)). On the other hand, when $\protect\omega _{\protect\phi _{1}}=1.1\protect\omega _{m}$, $\protect\omega _{\protect\phi _{2}}=0.9\protect\omega _{m}$, we see that the phase dispersion $\phi \left( t_{p}\right) $ is still zigzag with the change of $\Delta=\omega _{m}$, but the number of dips becomes three. Thus, in the case of $\protect\omega _{\protect\phi _{1}}$ is not equal to $\protect\omega _{\protect\phi _{2}}$, the fast and slow light effects can appear in more positions. This depends on the number of transparent windows. In addition, when the mechanical quality factor is very small, we find that the dip number decreases by one (see the dark blue curve in the last subgraph (d)). The physical mechanism behind this is that the thermal reservoir in contact with the vibrator will destroy the interference process, which is similar to the mechanism in Fig.  (\ref{Fig5}). Next we will explore the influence of these physical factors on fast and slow light.

In Fig. (\ref{Fig8}), we draw the  density plot of the group delay $\tau _{g}$ versus the  coupling field power and the angular momentum quantum number for different values of mechanical quality factors. We can clearly see the dependence of group delay $\tau _{g}$ on coupling power and the angular momentum quantum number.More specifically, in  Figs. \ref{Fig8}(a), \ref{Fig8}(b) and \ref{Fig8}(c), we find that fixing the coupling field power $P$ (the angular momentum quantum number) and adjusting the angular momentum quantum number $L$ (the  coupling field power) can change group delay $\tau _{g}$ from positive to negative, or from negative to positive, that is, adjusting the angular momentum can achieve the adjustment of fast and slow light. This provides a new degree of freedom for the adjustment of fast and slow light, which is not available in the traditional  Fabry-P\'erot cavity. Interestingly, as illustrated in Fig. \ref{Fig8}(a) that the group delay can even reach $-400$ $\mu$s and $+400$ $\mu$s, which is a typical ultrafast and ultraslow lights effect. However, with the increase of the mechanical quality factor, we can clearly see that the fast and slow light effects are not significant, and even only weak fast light effects are shown in  Fig. \ref{Fig8}(d). The physical reason behind this is that the thermal reservoir destroyed the OMIT effect caused by the RM1. Hence, we can speculate that if set $\Delta=0.9\omega _{m}$, modifying $Q_{2}$ but keep  $Q_{1}$ unchanged has similar results.

\begin{figure}
\centering
\includegraphics[width=0.5\textwidth]{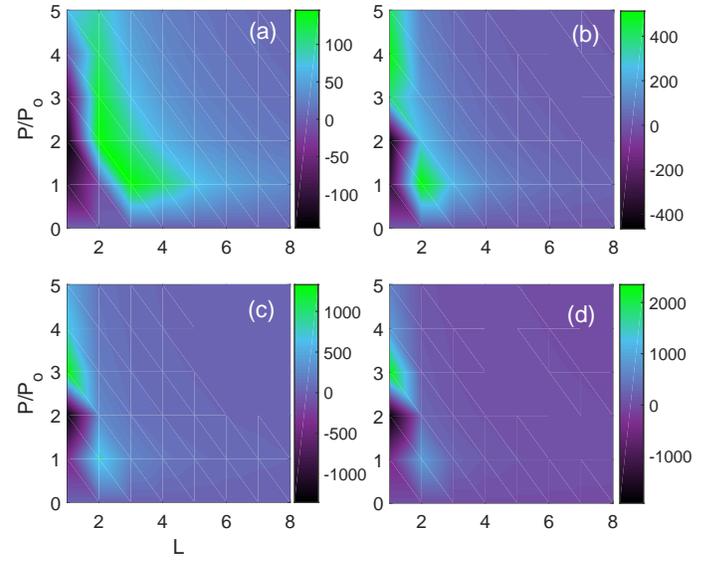}
\caption{The density plot of the group delay $\tau _{g}$ (in $\mu$s) versus the  coupling field power and the angular momentum quantum number for the different cavity field attenuation. (a) $\kappa =2\pi \times 35$ MHz, (b) $\kappa =2\pi \times 25$ MHz, (c) $\kappa =2\pi \times 15$ MHz, (d)$\kappa =2\pi \times 5$ MHz. Other parameters are the same as those in Fig. (\ref{Fig8}).} \label{Fig9}
\end{figure}

On the other hand, we also explore the influence of cavity field attenuation on the group delay $\tau _{g}$.  As expected, the group delay is closely related to the attenuation of the cavity field. Specifically, with the increase of cavity field attenuation, the effect of fast and slow light becomes less significant. By Comparing Figs. \ref{Fig9}(a) and \ref{Fig9}(d), we found that the group delay can differ to some extent. Interestingly, unlike the attenuation of the oscillator, even if the cavity field attenuation is large, we can still get ultrafast and ultraslow light (compare Fig. \ref{Fig8}(d) and Fig. \ref{Fig9}(a)).\;This advantage comes from the ``the double-OMIT phenomenon is robust against cavity field attenuation" mentioned in the description of the Fig. (\ref{Fig6}). Note that although the cavity field attenuation is not very destructive to the OMIT phenomenon in the currently studied system (see Fig. (\ref{Fig6})), and when it increases then it may enhance the dip width of the phase dispersion $\phi \left( t_{p}\right) $ (no longer sharp), which causes $\phi \left( t_{p}\right)$ to change slowly with $\omega _{p}$ near the dip (see Eq. \eqref{e9}). Therefore, the cavity field attenuation leads to the weakening of fast and slow light effects, and the mechanism and Fig. (\ref{Fig8}) are not completely consistent.

\section{Conclusion}
We have presented a two rotating mirror single mode Laguerre-Gaussian cavity optomechanical system. We have accomplished the twofold optomechanically induced transparency of the Gaussian probe field in rotational degrees of freedom. We have shown that the central absorption peak can become narrower and the two OMIT dips can become wider if the strength of the coupling field is increased. Furthermore, when the angular frequencies of the two rotating mirrors coincide, the system exhibits properties similar to a normal optomechanical Fabry-P\'erot cavity system in which only one mirror is moving. We also demonstrate the impact of the orbital angular momentum of the coupling field and rotating cavity decay on the output field. This shows that as the orbital angular momentum increases, the destructive interference increases in intensity and the transparency windows of the output probe field's absorption grow larger. This is consistent with the larger orbital angular momentum being associated with a larger optorotational coupling. We have shown that it is possible to switch between ultrafast and ultraslow light by altering the angular momentum, which is not possible in conventional optomechanical systems. The study of our paper has bright
prospects for the potential application of the rotating cavity optomechanics.
\begin{acknowledgements}
NA acknowledges the support of postdoctoral fund of the Zhejiang Normal University under Grant No.~ZC304021937. GX acknowledges support by the National Natural Science Foundation of China under Grants Nos.\ 11835011 and \ 12174346.
\end{acknowledgements}

\end{document}